\documentclass[12pt]{article}
\usepackage{amssymb,amsmath,epsfig}

\begin{document}

\title{\bf Generalized Teleparallel Gravity Via Some Scalar Field Dark Energy Models}
\author{M. Sharif \thanks {msharif.math@pu.edu.pk} and
Shamaila Rani\thanks {shamailatoor.math@yahoo.com}\\
Department of Mathematics, University of the Punjab,\\
Quaid-e-Azam Campus, Lahore-54590, Pakistan.}

\date{}

\maketitle
\begin{abstract}
We consider generalized teleparallel gravity in the flat FRW
universe with a viable power-law $f(T)$ model. We construct its
equation of state and deceleration parameters which give accelerated
expansion of the universe in quintessence era for the obtained scale
factor. Further, we develop correspondence of $f(T)$ model with
scalar field models such as, quintessence, tachyon, K-essence and
dilaton. The dynamics of scalar field as well as scalar potential of
these models indicate the expansion of the universe with
acceleration in the $f(T)$ gravity scenario.
\end{abstract}

\section{Introduction}

There are growing evidences of dark energy (DE) responsible for the
present expanding universe with an acceleration over the last few
years. Its confirmation is made by type Ia supernovae (Perlmutter et
al. 1999), galaxy redshift surveys (Fedeli et al. 2009), cosmic
microwave background radiation (CMBR) data (Caldwell and Doran 2004;
Huang et al. 2006; Keum 2007) and large scale structure (Koivisto
and Mota 2006; Daniel 2008). The standard cosmology has been
remarkably successful but there remain some serious unresolved
issues including the search for the best DE candidate. The origin
and nature of DE is still unknown except some particular ranges of
the equation of state (EoS) parameter $\omega$. In the absence of
any solid argument in favor of DE candidate, a variety of models
have been investigated.

Scalar field models are one of the proposed scenarios for DE. The
mechanism of these models suggests that a scalar field $(\phi)$
provides energy with negative pressure, leading to decrease a proper
potential of the field. A great number of scalar field DE models
have been studied so far, including quintessence with dominating
potential (Huang et al. 2006), K-essence with non-standard kinetic
term (Armendariz-Picon et al. 2000, 2001), tachyon having negative
squared mass (Sen 2002; Padmanabhan 2002), phantom keeping negative
energy (Nojiri and Odintsov 2003a, 2003b), ghost condensate with no
potential (Arkani-Hamed et al. 2004; Piazza and Tsujikawa 2004),
quintom (Guo et al. 2005; Zhang 2005; Setare 2006) and dilaton with
high energy particles (Copeland et al. 2006). There are many
attempts to reconstruct potential and scalar fields by establishing
a connection between different DE models with these scalar field
models.

Setare (2007a, 2007b, 2007c, 2007d, 2008) studied the correspondence
of HDE model with Chaplygin gas, interacting generalized Chaplygin
gas, interacting phantom scalar field and tachyon scalar field model
in general relativity. Ebrahimi and Sheykhi (2011) reconstructed the
power-law entropy-corrected HDE by correspondence with the above
mentioned scalar fields in non-flat evolving universe. Sharif and
Jawad (2012) have investigated interacting HDE with new IR cutoff to
develop correspondence with the scalar field models and discussed
the accelerated expansion of the universe. Granda and Oliveros
(2009) studied the correspondence between the quintessence, tachyon,
K-essence and dilaton energy density with HDE density by taking
event horizon as IR cutoff in the flat FRW universe. They
reconstructed potentials and dynamics for the scalar field models,
which describe accelerated expansion.

The $f(T)$ theory of gravity (Linder 2010) is the generalized form
of teleparallel gravity (Nashed 2002; Sharif and Amir 2007a, 2007b,
2008), which attracted many people to explore the accelerated
expansion of the universe. Its dynamics governs the torsion scalar
which takes part in this expansion. This theory uses
Weitzenb$\ddot{o}$ck connection which inherits only torsion and zero
curvature. There are many viable models (Bengochea and Ferraro 2009;
Wu and Yu 2010; Yang 2011; Ferraro and Fiorini 2011; Bamba 2011; Wei
et al. 2012) proposed in $f(T)$ theory to discuss DE era. Sharif and
Rani (2011) extended this work for Bianch type I universe and
discussed the accelerated expansion of the universe. They also
remarked that these models bear no equivalence at small scales.
Daouda et al. (2012) reconstructed the HDE model in this theory and
observed the crossing of phantom divide line. They also provided the
unification of dark matter and DE in this scenario.

In this paper, we establish correspondence between $f(T)$ gravity
and scalar field models in the flat FRW universe and discuss them
graphically. We obtain the EoS and deceleration parameters and
explore their behavior for the derived scale factor using a specific
$f(T)$ model. The paper is organized as follows: In section
\textbf{2}, the basic formalism of $f(T)$ and the field equations
are given. Section \textbf{3} explores  the EoS and deceleration
parameters for a particular $f(T)$ model. In section \textbf{4}, we
construct of $f(T)$ scalar field models. The last section contains
the summary of our results.

\section{The Generalized Teleparallel Gravity}

In this section, we introduce basic formalism of teleparallel as
well as its generalization $f(T)$ theory of gravity. The basic
ingredient in the structure of these theories is the vierbein field
$h_{a}(x^{\mu})$ (Nashed 2002; Sharif and Amir 2007a, 2007b, 2008)
which forms an orthonormal basis for the tangent space at each point
$x^\mu$ of the manifold. Here, the Latin alphabets
$(a,b,...=0,1,2,3)$ denote the tangent space indices and the
spacetime indices are represented by Greek alphabets
$(\mu,\nu,...=0,1,2,3)$. Each vector $h_a$ can be identified by its
components $h_\mu^a$ such that $h_a=h^\mu_a
\partial_\mu$. These tetrad are related to the metric tensor
$g_{\mu\nu}$ by the following relation
\begin{equation}\label{1}
g_{\mu\nu}=\eta_{ab}h_{\mu}^{a}h_{\nu}^{b},
\end{equation}
where $\eta_{ab}=diag(1,-1,-1,-1)$ is the Minkowski metric for the
tangent space and satisfy the following properties
\begin{equation}\label{2}
h^{a}_{\mu}h^{\mu}_{b}=\delta^{a}_{b},\quad
h^{a}_{\mu}h^{\nu}_{a}=\delta^{\nu}_{\mu}.
\end{equation}

The torsion scalar is given as
\begin{equation}\label{3}
T=S_{\rho}~^{\mu\nu}T^{\rho}~_{\mu\nu},
\end{equation}
where $S_{\rho}~^{\mu\nu}$ and torsion tensor $T^{\rho}~_{\mu\nu}$
are defined as follows
\begin{eqnarray}\label{4}
S_{\rho}~^{\mu\nu}&=&\frac{1}{2}(K^{\mu\nu}~_{\rho}
+\delta^{\mu}_{\rho}T^{\theta\nu}~_{\theta}-\delta^{\nu}_{\rho}T^{\theta\mu}~_{\theta}),\\
\label{5}T^{\lambda}~_{\mu\nu}&=&\Gamma^{\lambda}~_{\nu\mu}-
\Gamma^{\lambda}~_{\mu\nu}=h^{\lambda}_{a}
(\partial_{\nu}h^{a}_{\mu}-\partial_{\mu}h^{a}_{\nu}),
\end{eqnarray}
and $K^{\mu\nu}~_{\rho}=-\frac{1}{2}(T^{\mu\nu}~_{\rho}
-T^{\nu\mu}~_{\rho}-T_{\rho}~^{\mu\nu})$ is the contorsion tensor.
The action for $f(T)$ gravity is given by (Bengochea and Ferraro
2009; Wu and Yu 2010; Yang 2011; Ferraro and Fiorini 2011; Bamba
2011; Wei et al. 2012)
\begin{equation}\label{6*}
S=\frac{1}{2\kappa^2}\int d^{4}x[ef(T)+L_m],
\end{equation}
where $e=\sqrt{-g},~\kappa^{2}=8\pi G,~G$ is the gravitational
constant and $L_m$ is the Lagrangian density of matter inside the
universe. Here $f(T)$ is the general differentiable function of $T$.
The corresponding field equations are obtained by varying this
action with respect to vierbein as
\begin{equation}\label{6}
[e^{-1}\partial_{\mu}(eS_{a}~^{\mu\nu})
+h^{\lambda}_{a}T^{\rho}~_{\mu\lambda}S_{\rho}~^{\nu\mu}]f_{T}
+S_{a}~^{\mu\nu}\partial_{\mu}(T)
f_{TT}+\frac{1}{4}h^{\nu}_{a}f=\frac{1}{2}\kappa^{2}h^{\rho}_{a}T^{\nu}_{\rho},
\end{equation}
where $f_T,~f_{TT}$ stand for the first and second derivatives with
respect to $T$ and $T^{\nu}_{\rho}$ is the energy-momentum tensor of
the perfect fluid.

\section{Some Cosmological Parameters}

Here we discuss accelerated expansion of the universe through EoS
and deceleration parameters for the flat FRW universe described by
\begin{equation}\label{7}
ds^{2}=dt^{2}-a^{2}(t)(dx^{2}+dy^{2}+dz^{2}),
\end{equation}
where $a$ is the time dependent scale factor. The corresponding
tetrad components are $h^{a}_{\mu}=diag(1,a,a,a),$ which satisfy
Eq.(\ref{2}). Substituting these tetrad components in Eq.(\ref{3}),
the torsion scalar becomes $T=-6H^2$. Using these equations for
$a=0=\nu$ and $a=1=\nu$ in Eq.(\ref{6}), we obtain the following
modified Friedmann equations
\begin{eqnarray}\label{8}
12H^2f_T+f&=&2\kappa^{2}\rho,\\\label{9}
48H^2\dot{H}f_{TT}-(12H^2+4\dot{H})f_T-f&=&2\kappa^{2}p,
\end{eqnarray}
where $\rho$ and $p$ are the total energy density and pressure of
the universe and $H(=\dot{a}/a)$ is the Hubble parameter with dot
representing the time derivative. We assume here a pressureless
universe, i.e., $p_m=0$ and $\kappa^2=1$ for the sake of simplicity.
The above equations can be rewritten as
\begin{eqnarray}\label{10}
H^2&=&\frac{1}{3}(\rho_m+\rho_T),\\\label{11}
\dot{H}+H^2&=&-\frac{1}{6}(\rho_m+\rho_T+3p_T).
\end{eqnarray}
Here the subscripts $m$ and $T$ denote the matter and torsion
contributions of energy density and pressure, and are given as
follows
\begin{eqnarray}\label{12}
\rho_T&=&\frac{1}{2}(-12H^2f_T-f+6H^2),\\\label{13}
p_T&=&-\frac{1}{2}(48\dot{H}H^2f_{TT}-(12H^2+4\dot{H})f_T-f+6H^2+4\dot{H}),
\end{eqnarray}
and satisfying the energy conservation equation
\begin{equation}\label{14}
\dot{\rho}_T+3H(1+\omega_T)\rho_T=0.
\end{equation}
Noted that by taking $f(T)=T$ in the modified Friedmann equations
(\ref{10}) and (\ref{11}) which yield $\rho_{T}=0=p_{T}$ and the
resulting equations become the usual Friedmann equations in general
relativity. Also, the energy conservation equation and its solution
for dust matter are given by
\begin{equation}\label{15}
\dot{\rho}_m+3H\rho_m=0,\quad \rho_{m}=\rho_{0m}a^{-3},
\end{equation}
where $\rho_{0m}$ is an arbitrary constant. Using Eqs.(\ref{12}) in
(\ref{14}), the EoS parameter for $f(T)$ gravity is obtained as
\begin{equation}\label{18}
\omega_T=-1+\frac{4\dot{H}(12H^2f_{TT}-f_{T}+1)}{12H^2f_T+f-6H^2}.
\end{equation}
We take the following viable power law $f(T)$ model (Wu and Yu 2010;
Wei et al. 2012) to discuss the cosmic evolution of the universe,
\begin{equation}\label{19}
f(T)=\mu(-T)^n;\quad
\mu=(6H_0^2)^{1-n}\left(\frac{\Omega_{m0}}{2n-1}\right),
\end{equation}
where $n$ is a real constant and $\Omega_{m0}$ is the dimensionless
matter energy density. The index $0$ refers to the present values of
the corresponding quantities. Substituting this model along with
Eq.(\ref{12}) in (\ref{10}), we obtain the scale factor as follows
\begin{equation}\label{20}
a(t)=\left(\frac{3}{2n}\right)^{\frac{2n}{3}}\left(\frac{2\rho_{0m}}
{6^n\mu(1-2n)}\right)^{\frac{1}{3}}t^{\frac{2n}{3}}.
\end{equation}
We may compare this scale factor with the exact power law form of
scale factor used in literature as $a(t)=a_0t^m$ where $m>0$ and
$a_0$ is a constant (Sadjadi 2006; Nojiri and Odintsov 2006). For
$m>1$, it shows accelerating regime, while $0<m<1$ corresponds to
the decelerating era of the universe. Similarly, the scale factor
(\ref{20}) describes the expansion with acceleration and
deceleration of the universe for $n>3/2$ and $0<n<3/2$ respectively.
The expansion history of the universe has experienced a rapid
expansion and power law like decelerating as well as accelerating
phases. Thus it would be interesting to study these kinds of scale
factors in the modified gravity models. Inserting Eqs.(\ref{19}) and
(\ref{20}) in (\ref{18}), we obtain the following form of EoS
parameter
\begin{equation}\label{21}
\omega_T=\frac{(1-n)(3t^2)^{n-1}}{n[\mu(2n-1)(8n^2)^{n-1}+(3t^2)^{n-1}]}.
\end{equation}
\begin{figure} \centering
\epsfig{file=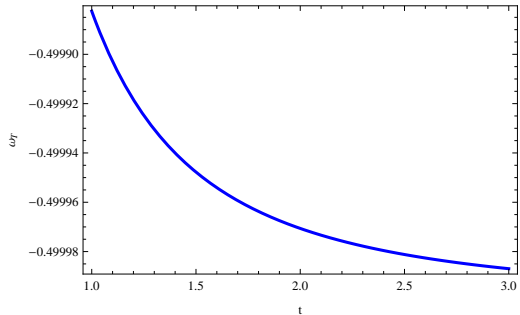,width=.50\linewidth}\caption{Plot of $\omega_T$
versus $t$.}
\end{figure}

The graphical behavior of $\omega_T$ versus time is shown in Figure
\textbf{1}. We use the current values $\Omega_{m0}=0.272,~H_0=74.2$
and assume the real constant as $n=2$. We examine that the EoS
parameter represents the quintessence region $(-1<\omega\leq-1/3)$
of the expanding universe. For the scale factor (\ref{20}), the
deceleration parameter is given by
\begin{equation}\label{a}
q_T=-1-\frac{\dot{H}}{H^2}=-1+\frac{3}{2n}.
\end{equation}
The negative behavior of this parameter is achieved for $n>3/2$,
which represents the accelerated expansion of the universe, whereas
$n\leq3/2$ corresponds to the decelerated phase of the universe.

\section{$f(T)$ Scalar Field Models}

In this section, we develop the correspondence of the $f(T)$ model
with some scalar field models like quintessence, tachyon, K-essence
and dilaton field models (Setare 2007c; van der Plas 2008) in the
flat universe.

\subsection{$f(T)$ Quintessence Model}

The dynamics of quintessence scalar field is governed by an ordinary
scalar field which slowly rolls down the potential. Slow-roll is the
condition in which kinetic energy of the system is less than the
potential energy, yielding the negative pressure. Its EoS parameter
describes accelerated expansion of the universe in the interval
$-1\leq\omega_q<-1/3$. The energy density and pressure of
quintessence scalar field are given by (Copeland et al. 2006)
\begin{eqnarray*}
\rho_{q}=\frac{1}{2}\dot{\phi}^{2}+V(\phi),\quad
p_{q}=\frac{1}{2}\dot{\phi}^{2}-V(\phi),
\end{eqnarray*}
where $\dot{\phi}^{2}$ and $V(\phi)$ are the kinetic energy and
scalar potential respectively. These take the form
\begin{figure} \centering
\epsfig{file=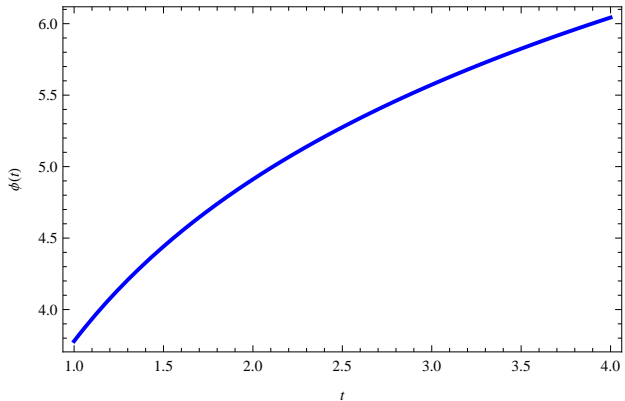,width=.60\linewidth}\caption{Plot of $\phi$
versus $t$ for quintessence model.}
\end{figure}
\begin{eqnarray}\label{22}
\dot{\phi}^{2}=\rho_q(1+\omega_q),\quad
V(\phi)=\frac{1}{2}\rho_q(1-\omega_q),
\end{eqnarray}
where the subscript $q$ represents the quantities corresponding to
quintessence model. In order to apply the correspondence, we equate
$\rho_T=\rho_q$ and $\omega_T=\omega_q$ and use Eqs.(\ref{12}) and
(\ref{21}) in (\ref{22}), it follows that
\begin{eqnarray}\label{23}
\dot{\phi}&=&\sqrt{\frac{4n}{3t^2}}\left[\mu(2n-1)
\left(\frac{8n^2}{3t^2}\right)^{n-1}+1\right]^{\frac{1}{2}},\\\label{24}
V(t)&=&\frac{2n}{3t^2}\left[n\left\{\mu(2n-1)
\left(\frac{8n^2}{3t^2}\right)^{n-1}+2\right\}-1\right].
\end{eqnarray}
\begin{figure} \centering
\epsfig{file=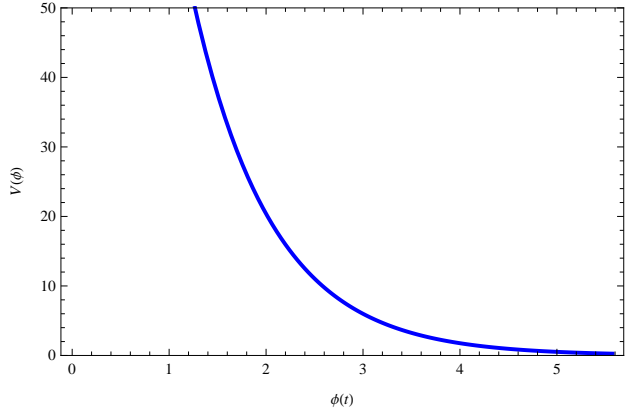,width=.60\linewidth} \caption{Plot of $V$
versus $\phi$ for quintessence model.}
\end{figure}

The analytical solution of Eq.(\ref{23}) is difficult to determine,
so we search for numerical solutions with the initial condition
$\phi(0)=0$. Figure \textbf{2} shows the plot of $\phi$ versus $t$
for the same parameters as in the previous section. This represents
an increasing behavior of the scalar field as time elapses and
results the decrement in kinetic energy of the potential. Figure
\textbf{3} indicates that the scalar potential is decreasing with
respect to scalar field up to a certain range and then becomes zero.
It corresponds to the scaling solution $(V(\phi)\propto\phi^{-1})$
(Copeland et al. 2006; Sharif and Jawad 2012) which represents the
accelerated expansion of the universe. The slow-roll condition is
satisfied for certain range as the kinetic energy is less than the
potential energy. Hence, the scalar field $\phi$ slowly rolls down
the scalar potential in the $f(T)$ quintessence model (van der Plas
2008).

\subsection{$f(T)$ Tachyon Model}

The tachyon scalar field model has the energy density and pressure
as (Copeland et al. 2006; Setare 2007c)
\begin{eqnarray}\label{25}
\rho_{t}=\frac{V(\phi)}{\sqrt{1-\dot{\phi}^{2}}},\quad
p_{t}=-V(\phi)\sqrt{1-\dot{\phi}^{2}},
\end{eqnarray}
leading to the EoS parameter
\begin{equation}\label{26}
\omega_{t}=\dot{\phi}^{2}-1.
\end{equation}
This equation indicates a universe dominated by cosmological
constant in the limit of vanishing kinetic energy. The
correspondence of $f(T)$ energy density and EoS parameter with
tachyon model yields
\begin{eqnarray}\label{27*}
\dot{\phi}&=&\left[\frac{1-n}{n\{\mu(2n-1)
(\frac{8n^2}{3t^2})^{n-1}+1\}}+1\right]^{\frac{1}{2}},\\\label{27}
V(t)&=&\frac{4n^2}{3t^2}\sqrt{1-\frac{1}{n}}\left[\mu(2n-1)
\left(\frac{8n^2}{3t^2}\right)^{n-1}+1\right]^{\frac{1}{2}}.
\end{eqnarray}
\begin{figure} \centering
\epsfig{file=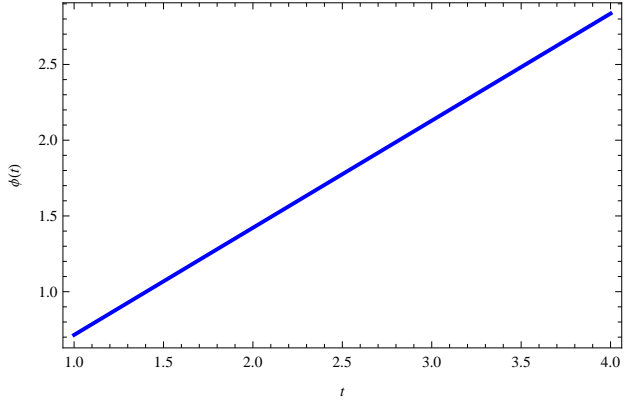,width=.60\linewidth} \caption{Plot of $\phi$
versus $t$ for tachyon model.}
\end{figure}
\begin{figure} \centering
\epsfig{file=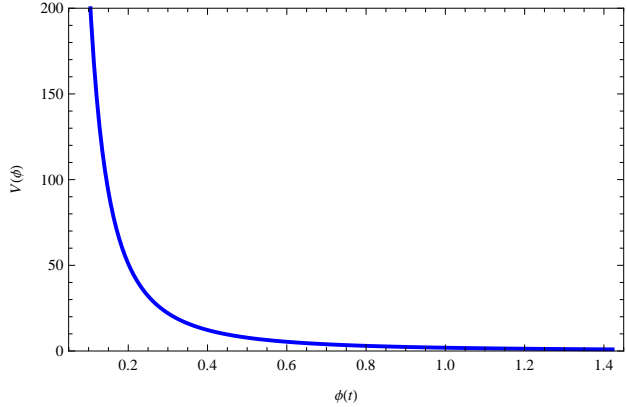,width=.60\linewidth} \caption{Plot of
$V(\phi)$ versus $\phi(t)$ for tachyon model.}
\end{figure}

Figures \textbf{4} and \textbf{5} show the evolution trajectories of
scalar field and potential versus $t$ and $\phi$ respectively. The
scalar field indicates the increasing behavior with direct
proportional to time (Sharif and Jawad 2012) leads to the continuous
expansion. Figure \textbf{5} shows the same behavior of potential as
for the previous model (Figure \textbf{3}). However the $f(T)$
tachyon potential shows the expansion of the universe as it rolls
down to its minimum value and corresponds to the scaling solution
(Copeland et al. 2006) for a small region as compared to the $f(T)$
quintessence model. It is noted here that interacting new
holographic tachyon model also behaves like scaling models (Sharif
and Jawad).

\subsection{$f(T)$ K-essence Model}

The K-essence scalar model (Copeland et al. 2006) gives the
accelerated expansion of the universe with the help of kinetic
energy $X$ and its modified forms. For FRW universe, kinetic energy
takes the form $X=\frac{1}{2}~\dot{\phi}^2$. Also, the energy
density and pressure are
\begin{eqnarray}\label{28}
\rho_{k}=V(\phi)(-X+3X^{2}),\quad p_{k}=V(\phi)(-X+X^{2}).
\end{eqnarray}
The corresponding EoS parameter is
\begin{equation}\label{29}
\omega_{k}=\frac{1-X}{1-3X}.
\end{equation}
This shows accelerating expansion of the universe with a specific
range of $X$. For $X=1/2$, EoS parameter behaves like cosmological
constant whereas acceleration boundary $\omega_k=-1/3$ is obtained
for $X=2/3$. Hence, the expanding universe with acceleration
corresponds to the interval $1/2\leq X<2/3$. Equating
$\rho_{k}=\rho_{T}$ and $\omega_{k}=\omega_{T}$ for the
correspondence, we obtain
\begin{equation}\label{30}
X=1+\frac{2(1-n)}{\left[n\left\{\mu(2n-1)
\left(\frac{8n^2}{3t^2}\right)^{n-1}+4\right\}-3\right]}.
\end{equation}
\begin{figure} \centering
\epsfig{file=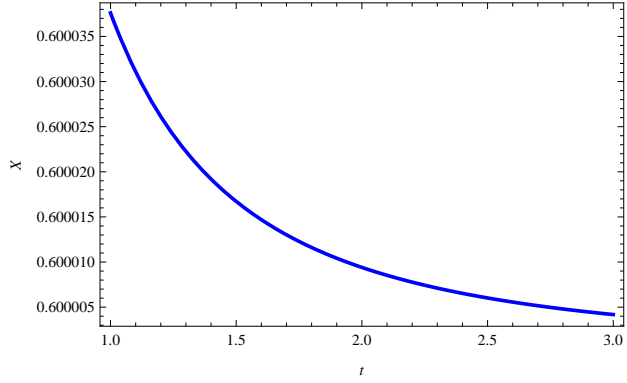,width=.60\linewidth} \caption{Plot of $X$
versus $t$ for K-essence model.}
\end{figure}
\begin{figure} \centering
\epsfig{file=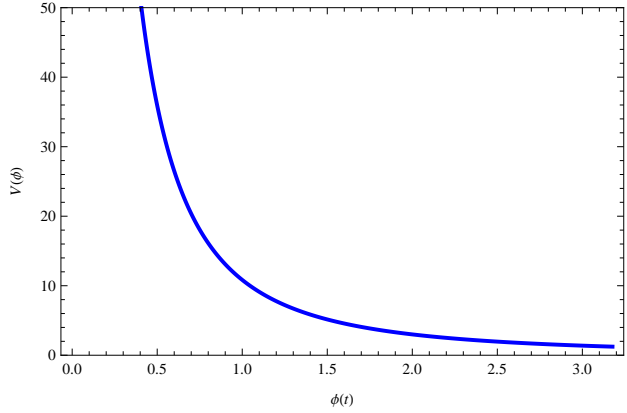,width=.60\linewidth} \caption{Plot of
$V(\phi)$ versus $\phi$ for K-essence model.}
\end{figure}
\begin{eqnarray}\nonumber
V(t)&=&\frac{4n^2}{3t^2}\left[\mu(2n-1)
\left(\frac{8n^2}{3t^2}\right)^{n-1}+1\right]\left[\frac{10(1-n)}
{n\left\{\mu(2n-1)(\frac{8n^2}{3t^2})^{n-1}+4\right\}-3}\right.
\\\label{31}&+&\left.12\left(\frac{(1-n)^2}{n\left\{\mu(2n-1)
(\frac{8n^2}{3t^2})^{n-1}+4\right\}-3}\right)^2+2\right]^{-1}.
\end{eqnarray}

The plot of $X$ versus $t$ shows consistent results within the
interval as shown in Figure \textbf{6}. Figure \textbf{7} represents
the same behavior of $V(\phi)$ as $f(T)$ quintessence and new
holographic tachyon model (Sharif and Jawad 2012) inherit. Also, the
relation $X=\frac{1}{2}\dot{\phi}^2$ yields
\begin{eqnarray}\label{32}
\dot{\phi}=\left(2+\frac{4(1-n)}{n\left\{\mu(2n-1)
\left(\frac{8n^2}{3t^2}\right)^{n-1}+4\right\}-3}\right)^{\frac{1}{2}}.
\end{eqnarray}
Its plot versus $t$ is given in Figure \textbf{8} which shows
increasing behavior (Sharif and Jawad 2012) of scalar field with
direct proportionality. It represents the continuous expansion of
the universe.
\begin{figure} \centering
\epsfig{file=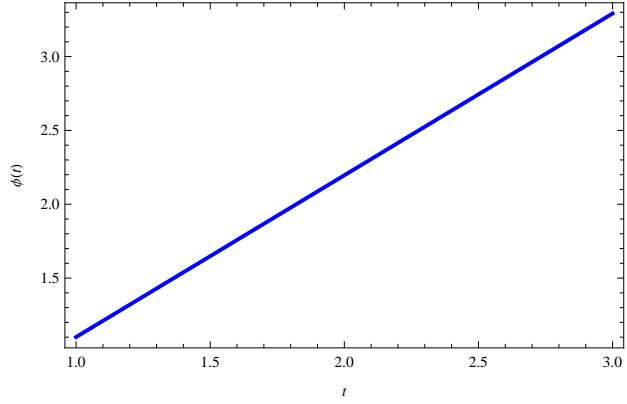,width=.60\linewidth} \caption{Plot of $\phi$
versus $t$ for K-essence model.}
\end{figure}

\subsection{$f(T)$ Dilaton Model}

The pressure of the dilaton scalar field model is given by (Piazza
and Tsujikawa 2004)
\begin{equation}\label{33}
p_d=-X+b_{1}e^{b_2\phi}X^2,
\end{equation}
where $b_{1}$ and $b_{2}$ are positive constants and
$2X=\dot{\phi}^2$. It is explained by a general four dimensional
effective low-energy string action. Its dynamics is governed by
negative kinetic term and higher-order derivative terms of
$\dot{\phi}$ to stable the system. The energy density of dilaton
model is
\begin{eqnarray}\label{34}
\rho_{d}&=&-X+3b_{1}e^{b_{2}\phi}X^{2},
\end{eqnarray}
Dividing Eq.(\ref{33}) by (\ref{34}), we obtain EoS parameter as
\begin{equation}\label{35}
\omega_{d}=\frac{1-b_{1}e^{b_{2}\phi}X}{1-3b_{1}e^{b_{2}\phi}X}.
\end{equation}
It meets the universe in accelerated expansion for the bound
$(\frac{20}{3},\frac{40}{3})$ of $e^{b_{2}\phi}X$. Replacing
$\omega_d$ by $\omega_T$ for $f(T)$ dilaton model, it follows that
\begin{eqnarray}\label{36}
e^{b_{2}\phi}X&=&\frac{1}{b_1}\left[1+\frac{2(1-n)}{n\left\{\mu(2n-1)
\left(\frac{8n^2}{3t^2}\right)^{n-1}+4\right\}-3}\right].
\end{eqnarray}
\begin{figure} \centering
\epsfig{file=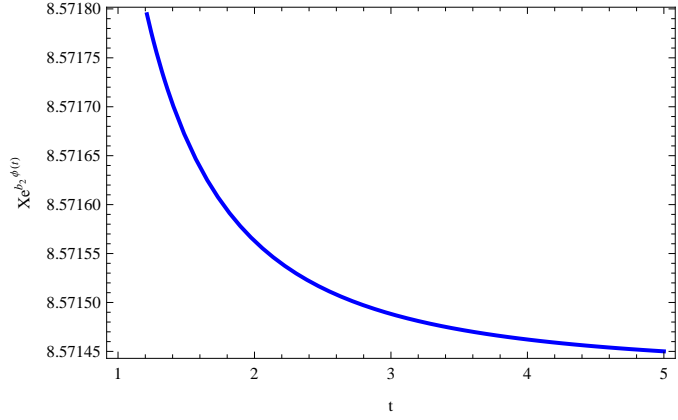,width=.65\linewidth} \caption{Plot of
$e^{b_{2}\phi}X$ versus $t$ for dilaton field.}
\end{figure}
\begin{figure} \centering
\epsfig{file=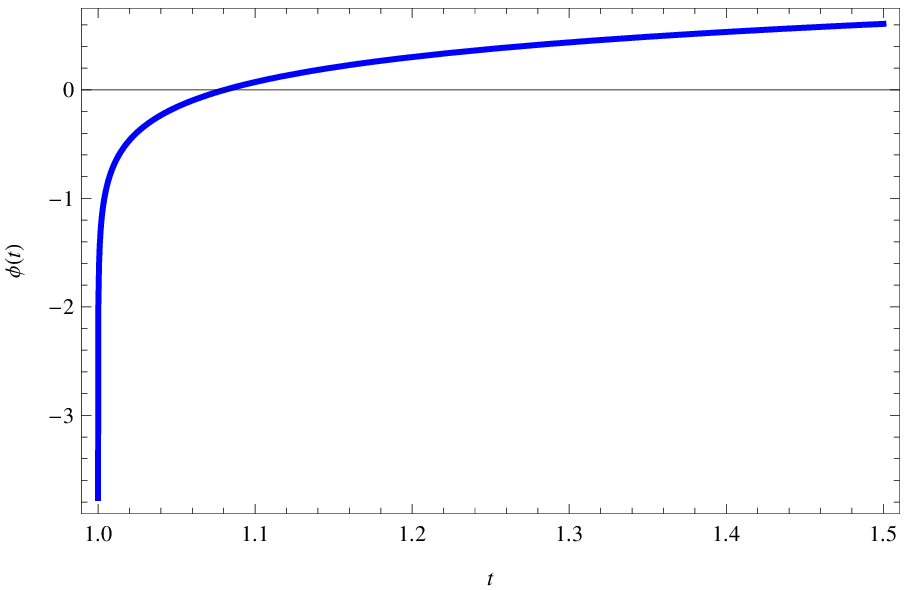,width=.60\linewidth} \caption{Plot of $\phi$
versus $t$ for dilaton field.}
\end{figure}

Figure \textbf{9} shows the graph of $e^{b_{2}\phi}X$ versus $t$
with $b_1=0.07,~b_2=6$. It satisfies the criteria for accelerated
expansion of the universe. Also, the solution of the above equation
is as follows
\begin{equation}\label{37}
\phi(t)=\frac{2}{b_{2}}\ln\left[\frac{b_2}{\sqrt{2b_1}}\int^{t}_{0}\left[1+\frac{2(1-n)}{n\left\{\mu(2n-1)
\left(\frac{8n^2}{3t^2}\right)^{n-1}+4\right\}-3}\right]^{\frac{1}{2}}dt\right].
\end{equation}
Figure \textbf{10} represents the cosmic evolution of scalar field.
Initially, it bears increasing negative values, however it becomes
positive after a small interval of time and shows flatness with the
passage of time. These types of solutions are scaling solutions due
to the relation $\phi(t)\propto \ln t$ (Sharif and Jawad 2012).

\section{Concluding Remarks}

The connection of scalar field models to different DE models has
gained a lot of interest due to its role in discussing accelerated
expansion of the universe. In this context, we have considered the
framework of $f(T)$ gravity to connect with scalar field DE models
such as, quintessence, tachyon, K-essence and dilaton. We have
discussed these models graphically by taking a viable power law
$f(T)$ model. We have derived the scale factor from the first
modified Friedmann equation in terms of power law form. Also, we
have checked the behavior of evolution trajectory of EoS parameter
and deceleration parameter of the model. The results of the paper
are summarized as follows.

The EoS parameter indicates the quintessence era of the DE dominated
universe whereas the deceleration parameter corresponds to this era
for the real constant $n>3/2$ of the model. We have provided a
correspondence between $f(T)$ model and some scalar field models to
analyze the accelerated expansion of the universe. The scalar field
and potential are studied graphically with respect to time. These
correspondences give
\begin{enumerate}
\item The scalar field graph of $f(T)$ quintessence model
represents increasing behavior while potential versus $\phi$
indicates scaling solution, leading to the accelerated expansion of
the universe. The slow-roll condition is satisfied up to a certain
region of scalar field in this case.
\item The plots of $\phi$ and $V$ of $f(T)$ tachyon model show the same
behavior as for the previous model but for smaller region with
scaling type solution.
\item In the $f(T)$ K-essence model, kinetic
energy shows DE region and the scalar field slowly rolls down the
potential which is directly proportional to the time. It indicates
an ever expanding universe.
\item For the correspondence between $f(T)$ and dilaton
model, the $e^{b_{2}\phi}X$ indicates the expansion of the universe.
Also, the scaling solution is obtained for its scalar field due to
$\phi(t)\propto \ln t$.
\end{enumerate}

We would like to mention here that mostly HDE and its modified
models are taken to make such type of correspondences (Granda and
Oliveros 2009; Ebrahimi and Sheykhi 2011; Sharif and Jawad 2012) to
discuss accelerated expansion of the universe. It is an effective
procedure for the correspondence of scalar field models with
different DE models which may help to comprehend the unknown DE
candidate. It is interesting to mention here that our results are
viable and consistent by comparing with those already available for
other dark energy models (Setare , 2007b, 2007c, 2007d, 2008; Granda
and Oliveros 2009; Ebrahimi and Sheykhi 2011; Sharif and Jawad
2012).

\end{document}